\documentclass[preprintnumbers,aps,twocolumn,floatfix,superscriptaddress]{revtex4-1}
\usepackage{eso-pic,calc}
\usepackage{graphicx}
\usepackage{amsmath, amsthm, amssymb}
\usepackage{epsfig}
\usepackage{bm}
\usepackage{color}

\usepackage[colorlinks=True]{hyperref}  
\hypersetup{
 	colorlinks=true,  
 	linkcolor=blue,  
 	citecolor=blue, 
 	filecolor=magenta,   
 	urlcolor=blue         
 }

\def\etl{$et~al.$~}

\begin{document}
\title{Fitness noise in the Bak-Sneppen evolution model in high dimensions}

\author{Rahul Chhimpa}
\affiliation{Department of Physics, Institute of Science,  Banaras Hindu University, Varanasi 221 005, India}

\author{Abha Singh}
\affiliation{Department of Physics, Institute of Science,  Banaras Hindu University, Varanasi 221 005, India}

\author{Avinash Chand Yadav\footnote{jnu.avinash@gmail.com}}
\affiliation{Department of Physics, Institute of Science,  Banaras Hindu University, Varanasi 221 005, India}

\begin{abstract}
{ We study the Bak-Sneppen evolution model on a regular hypercubic lattice in high dimensions. Recent work [\href{https://journals.aps.org/pre/abstract/10.1103/PhysRevE.108.044109}{Phys. Rev. E {\bf 108}, 044109  (2023)}] has shown the emergence of the $1/f^{\alpha}$ noise for the ``fitness'' observable with $\alpha \approx 1.2$ in one-dimension (1D) and $\alpha \approx 2$ for the random neighbor (mean-field) version of the model. We examine the temporal correlation of fitness in 2, 3, and 4 dimensions. As obtained by finite-size scaling, the spectral exponent tends to take the mean-field value at the upper critical dimension ${\rm D}_u = 4$, which is consistent with previous studies. Our approach provides an alternative way to understand the upper critical dimension of the model. We also show the local activity power spectra, which offer insight into return time statistics and the avalanche dimension.}
\end{abstract}

\maketitle

\section{Introduction}
Many natural systems evolve spontaneously towards a critical state, where observables show scaling features in space-time. Relevant to diverse contexts ranging from earthquakes~\cite{PhysRevLett.68.1244, PhysRevLett.88.228301, PhysRevLett.88.178501, PhysRevLett.114.088501, PhysRevLett.122.218501} to biological evolution~\cite{PhysRevLett.71.4083, PhysRevE.97.042123, PhysRevLett.132.098402}, the plausible mechanism that can offer an explanation of the scaling behaviors is self-organized criticality (SOC)~\cite{PhysRevLett.59.381, MARKOVIC201441, Pruessner_2012, Christensen_2005, Watkins2016}. To explain biological evolution, Bak and Sneppen (BS)~\cite{PhysRevLett.71.4083} introduced a minimalistic model based on the Darwinian evolutionary principles~\cite{darwin1859origin, koch201380}. An ecosystem consists of $L$ species arranged on a ring. Each species has a \emph{fitness}, initially drawn randomly and uniformly between 0 and 1. The \emph{least-fit} species undergoes extinction or mutation by replacing its fitness with the same distribution. The nearest neighboring species coevolved similarly to include interaction (food web dependence). The same mutation event occurs iteratively. The BS model has \emph{extremal} dynamics similar to the Robin Hood model~\cite{ZAITSEV1992411, PhysRevLett.75.1550} and the Sneppen interface model~\cite{PhysRevLett.69.3539}. 

Extensively studied, the BS model exhibits rich behavior, although the dynamics are simple.
A mutation event is active if at least one species has fitness below a threshold. The number of active mutation steps between two consecutive quiescent configurations is the extinction avalanche size. The size distribution satisfies a power law with an exponent $\tau \approx 1$ in the 1D BS model~\cite{PhysRevLett.71.4083, PhysRevE.53.414}.
Strikingly, the BS model displays \emph{punctuated equilibrium}~\cite{Gould1993}, i.e., evolution happens as an intermittent burst of activities separated by long stasis rather than gradual change. Unlike nonlinear dynamical systems, the BS model shows sensitivity (weak) to the initial condition~\cite{Tamarit1998}. To understand the role of \emph{dimensionality} for avalanches and return time statistics~\cite{PhysRevLett.73.2162}, the BS model has been studied on regular lattices in high dimensions, yielding an upper critical dimension of 4~\cite{PhysRevLett.84.2267, Han_2018} or 8~\cite{PhysRevLett.80.5746}. The BS model has also been examined for complex~\cite{PhysRevLett.81.2380, Moreno_2002} and adaptive networks~\cite{Garlaschelli2007,Caldarelli2008}. The BS model is robust even if the fitness noise is replaced by chaotic or quasiperiodic signals~\cite{PhysRevE.56.4876}. The $k$ traits generalization of the BS model is solvable in the limit $k\to \infty$, yielding $\tau=3/2$~\cite{PhysRevLett.76.348}. 

To understand the extent of SOC features for the BS model, several variants have been examined. (i) In the anisotropic BS (aBS) model~\cite{PhysRevE.58.7141}, only the nearest right neighbor of the least-fit site is updated. Although the avalanche size exponent $\tau$ remains the same, the avalanche dimension differs. Thus, the aBS model belongs to a different universality class. (ii) In the random-neighbor BS (rBS) model~\cite{PhysRevLett.71.4087}, one randomly selected site is updated along with the least-fit species. The rBS is solvable because the mean-field limit is applicable, resulting in $\tau = 3/2$~\cite{PhysRevLett.73.906}. (iii) Manna~\cite{PhysRevE.80.021132} introduced a stochastic BS (sBS) model where one of the nearest neighbors, selected randomly with equal probability, is updated. The sBS model on regular lattices belongs to the same universality class as the original BS model. SOC holds in the sBS model on complex networks with an average branching factor greater than 1. (iv) Although fitness is typically a continuous variable, two values are possible in the discrete BS model~\cite{Volkov2022}, where mutation takes place following the Bernoulli distribution. 

In previous studies, many noisy signals for the original BS model have been examined and found to exhibit the $1/f^{\alpha}$ noise~\cite{Yadav_2022} (also see references therein). The local activity (the recurrent activity at a given site) offers insight into the return time statistics. The local activity is $A(t) = 1$ if the site has the least-fit species at time $t$, and $A(t) = 0$ otherwise. Maslov \etl numerically predicted the spectral exponents $\alpha = 0.57$ in 1D and $\alpha = 0.32$ in two-dimension (2D) and suggested that the exponent satisfies a scaling relation $\alpha = 1-{\rm D}/\mathcal{D}$, where $\mathcal{D}$ is the avalanche dimension and D is the space dimension~\cite{PhysRevLett.73.2162, PhysRevE.53.414}. Abha \etl recently showed that while the exponent is $\alpha \approx 0.40$ for the aBS model in 1D, it is $\alpha = 1/2$ if the underlying motion for the least-fit site is a simple random walk~\cite{PhysRevE.108.044109}.

Daerden and Vanderzande~\cite{PhysRevE.53.4723} studied a different signal $n(t)$, the number of species below a threshold fitness. They reported that $n(t)$ shows $1/f^{\alpha}$ noise with $\alpha \approx 1$ for both the 1D BS model (supported only numerically) and the rBS model. Later, Davidsen and L\"uthje~\cite{PhysRevE.63.063101} commented that $n(t)$ in the rBS model is like an Ornstein-Uhlenbeck process (Brownian motion in a parabolic potential). This yields a Lorentzian spectrum with $\alpha \approx 2$ for the rBS model, and the exponent is $\alpha \approx 1.5$ in the 1D BS model. Abha \etl also provide numerical evidence that the exponent is $\alpha \approx 1.3$~\cite{PhysRevE.108.044109} in the 1D BS model, significantly differing from 1.

More recently, a different quantity, fitness noise, was examined in the 1D BS model. As shown numerically, the fitness exhibits the $1/f^{\alpha}$ noise with $\alpha \approx 1.2$~\cite{PhysRevE.108.044109}. However, the exponent changes to $\alpha \approx 2$ in the rBS model, which is the mean-field behavior. Although the exponent increases marginally with increasing the mutation zone (the number of interacting nearest neighbors) in 1D, it takes a value significantly smaller than the mean-field value.

\begin{figure}[b]
  \centering
       \scalebox{1}{\includegraphics{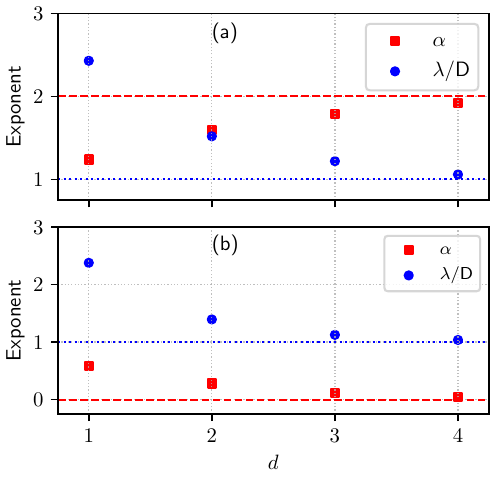}}
  \caption{The variation of critical exponents [the spectral $\alpha$ and the cutoff frequency $\lambda$ (discussed below)] as a function of space dimension ${\rm D}$ for (a) the fitness noise $\eta(t)$ and (b) the local activity $A(t)$. Dashed and dotted lines represent theoretical mean-field values for $\alpha$ and $\lambda/{\rm D}$, respectively. The ratio $\lambda/{\rm D}$ decreases with dimension and approaches 1 at ${\rm D}_u$. The spectral exponent changes until it reaches the mean-field value for ${\rm D}\ge {\rm D}_u$. The global fitness spectral exponent increases for $D\le {\rm D}_u$, while the local activity spectral exponent decreases.}
  \label{fig_exp_1}
\end{figure}

In the BS model on regular lattices, it remains unknown how the spectral exponent for the fitness noise changes with dimensionality. Although previous studies suggest the upper critical dimension for the BS model, ${\rm D}_u = 4$ or $8$, it is pertinent to determine when $\alpha$ approaches the mean-field value. In this paper, we undertake this issue and examine the BS model in high dimensions. Our study reveals that the spectral exponent for the fitness noise increases with dimensionality and approaches its mean-field value at ${\rm D}_u = 4$ [cf. Fig.~\ref{fig_exp_1}(a)].

We also examine the local activity power spectrum and find the spectrum to be flat at ${\rm D}_u = 4$, suggesting uncorrelated behavior [cf. Fig.~\ref{fig_exp_1}(b)]. 
We can also determine the avalanche dimension value $\mathcal{D}$ using the scaling relation. The estimated value of $\mathcal{D}$ is found to be consistent with previously determined values in an alternative way. As expected, the logarithmic correction at ${\rm D}_u = 4$ seems consistent with our results. 

The organization of the paper is as follows. Section~\ref{sec_2} presents the model definition and quantities of interest. In Sec.~~\ref{sec_3}, we discuss the finite-size scaling that is employed to determine critical exponents. In Sec.~~\ref{sec_4}, we show numerical results for the power spectra of fitness noise and local activity. Finally, Sec.~\ref{sec_5} summarizes the paper with a brief discussion.

\begin{figure}[t]
  \centering
       \scalebox{1}{\includegraphics{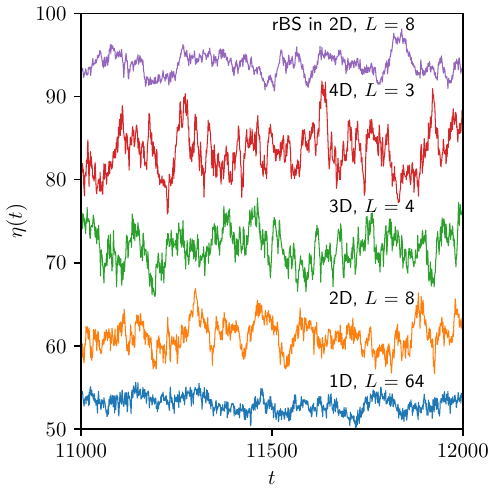}}
  \caption{A portion of the fitness noise $\eta(t)$ in different dimensions for the BS model after discarding $10^6$ transients. Each signal is shifted vertically by a constant for clear visibility. The top signal corresponds to the 2D rBS model, equivalently the mean-field behavior.}
  \label{fig_sig_1}
\end{figure}

\section{Model}{\label{sec_2}}
In ${\rm D}$-dimension, consider a regular hypercubic lattice with its linear extent $L$ and periodic boundary conditions. Each site represents a species characterized by fitness $\xi({\bf x}, t)$ (a measure of survivability). Initially, the fitness is drawn from a uniform probability density function in the unit interval. The dynamics include two steps. (i) At each time $t$, the least-fit species gets extinct or mutated. In place of that, a new species appears with a fitness value drawn from the same distribution. (ii) In turn, the 2D nearest neighbors are updated similarly, reflecting an interaction in space (coevolution). The dynamic is repeated ad infinitum. 

We examine two noisy signals: (i) the global fitness $\eta(t) = \sum_{\bf x}\xi({\bf x}, t)$ and (ii) the local activity $A({\bf x}, t)$ (that is, 1 if the site ${\bf x}$ is least fit at time $t$ and 0 otherwise). Employing the Monte Carlo simulation method, we generate the signals of length $N$ after discarding $10^6$ transients. Figure~\ref{fig_sig_1} shows typical global fitness noise for a fixed system size in different dimensions.

In the statistically stationary state, the fitness distribution is $p(\xi) = 1/(1-\xi_c)$ for $\xi_c < \xi <1$ with a critical fitness of $\xi_c \approx 2/3$ for the 1D BS model~\cite{PhysRevE.53.414}. Although the initial fitness distribution is uniform in the BS model, the SOC features also hold for exponential~\cite{PhysRevLett.75.1969} and power-law~\cite{PhysRevE.58.3993} distributed fitness. SOC is destroyed in the two limiting cases~\cite{PhysRevE.80.021132}. (i) If there is no interaction, $\xi_c \to 1$. (ii) If each node is linked to all other nodes (a complete network), then $\xi_c \to 0$. In the BS model on a scale-free network, the critical fitness tends to be 0 in the thermodynamic limit, while it remains finite in the sBS model. 
The least-fit site evolves in space-time as a L\'evy walk with a power-law distributed jump.

The occurrence of the least fit at any particular site can happen frequently. Typically, such a feature is characterized by power law return-time statistics. Two exponents, namely, the first return time $\tau_f$ and the all return time $\tau_a$, satisfy the scaling relation $\tau_f+\tau_{\rm all} = 2$ (recurrent) if $\tau_{\rm all} \leq1$ and $\tau_{\rm all} = \tau_f$ (transient) if $\tau_{\rm all}>1$~\cite{PhysRevLett.80.5746}. The mean-field values are $\tau_f = \tau_{\rm all} = 1$~\cite{PhysRevLett.73.2162} or 3/2~\cite{PhysRevLett.80.5746}. For a simple random walk, it is well known that $\tau_f = 3/2$ and $\tau_{\rm all} = 1/2$ in 1D and $\tau_f = \tau_{\rm all} = {\rm D}/2$ in ${\rm D}\ge 2$.

\section{Method}{\label{sec_3}}
To uncover the underlying temporal correlation of the noisy processes, we compute the power spectrum $S_{\eta}(f) =  \lim_{N\to \infty} \langle |\tilde{\eta} (f)|^2\rangle/N$, where the angular brackets $\langle \cdot \rangle$ denote ensemble average over $M$ independent realizations. We use a standard fast Fourier transform (FFT) algorithm to compute the Fourier transformation of the process $\tilde{\eta} (f)$ given as 
 \begin{equation}
\tilde{\eta} \left(f = \frac{k}{N}\right)=  \frac{1}{\sqrt{N}} \sum_{t=0}^{N-1}\eta(t) \exp\left( -j2\pi \frac{k}{N}t\right),\nonumber
\label{eq_ps_x1}
\end{equation}
with $k = 0, 1, 2, \cdots, N-1$.

Interestingly, the power spectrum exhibits scaling behavior of the $1/f^{\alpha}$ type in the frequency regime $L^{-\lambda}\ll f\ll 1/2$. One can employ the finite-size scaling method to determine the critical exponents $\alpha$ and $\lambda$. 
The variation of power spectra for different system sizes (as shown below) reveals that the power can show, in general, system size scaling features in both frequency regimes as
 \begin{equation}
S_{\eta}(f,L) \sim \begin{cases} L^a, ~~~~~~~~~~~~~~~~~~ f\ll L^{-\lambda},    \\L^{a-\lambda\alpha}/f^{\alpha}, ~~~~~~L^{-\lambda}\ll f \ll 1/2. 
\end{cases}
\label{eq_ps1}
\end{equation} 
Notice that one can easily obtain data collapse by plotting scaled power $S_{\eta}(f, L)/L^a$ in terms of reduced frequency $fL^{\lambda}$. 
It implies that Eq.~(\ref{eq_ps1}), as noted in previous studies~\cite{PhysRevE.104.064132, Kumar_2022, PhysRevE.108.044109}, satisfies a scaling ansatz, $S_{\eta}(f, L) \sim L^{a}H(fL^{\lambda}) $, where the scaling function $H(u)$ is constant for $u\ll1$ and $u^{-\alpha}$ for $u\gg1$. 
The total power also shows a system size scaling $P(L)\sim \int S_{\eta}(f, L) df \sim L^{a-\lambda}$. The system size scaling of the total power and the power in the low-frequency component can directly provide an estimate of the two critical exponents $a$ and $\lambda$. 

\begin{figure}[t]
  \centering
       \scalebox{1.0}{\includegraphics{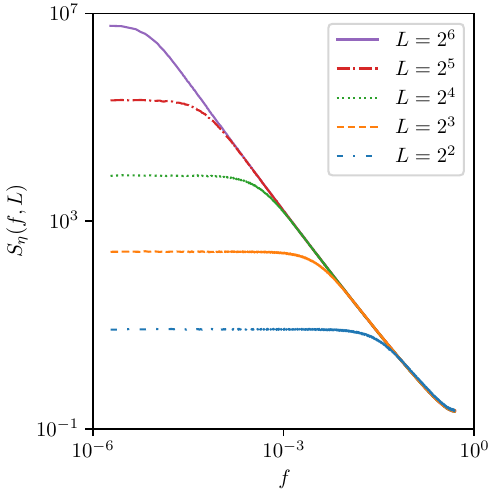}}
  \caption{The global fitness power spectra for the 2D BS model with different system sizes. The signal length is $N = 2^{20}$, and the ensemble averaging is performed over $10^4$ independent realizations of the signal. }
  \label{fig_2d_gf_1}
\end{figure}

\begin{figure}[t]
  \centering
       \scalebox{1.0}{\includegraphics{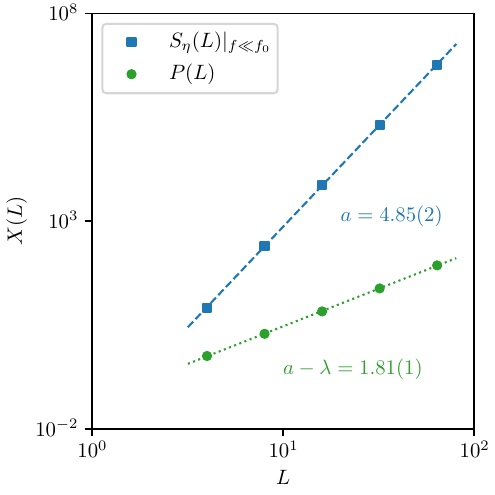}}
  \caption{For the 2D BS model, the system size variation of the global fitness power in the low frequency component and the total power. Straight lines are the least squares fit.}
  \label{fig_2d_gf_2}
\end{figure}

 \section{Simulation results}{\label{sec_4}} 
 \subsection{Global fitness}
\begin{figure}[t]
  \centering
       \scalebox{1.0}{\includegraphics{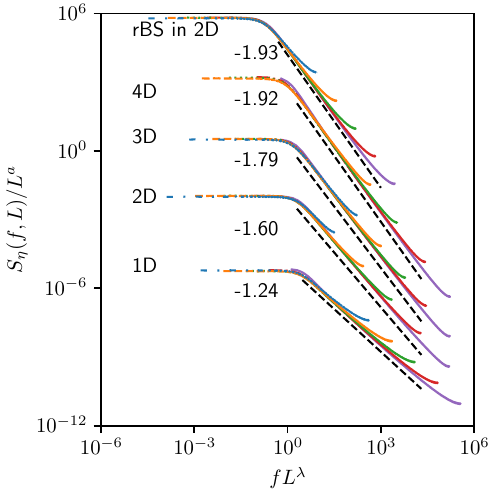}}
  \caption{The scaling function of global fitness power spectra for the BS model in different dimensions. For clarity, the curves are shifted vertically by a constant factor.}
  \label{fig_gf_1}
\end{figure}

 Figure~\ref{fig_2d_gf_1} shows the global fitness power spectra for different system sizes in the 2D BS model. The power remains constant for a frequency regime $f\ll L^{-\lambda}$ and varies as $1/f^{\alpha}$ for the non-trivial frequency regime $L^{-\lambda}\ll f \ll 1/2$. Since the power spectra do not show system size dependence in the non-trivial frequency regime, one can immediately verify from Eq.~(\ref{eq_ps1}) a scaling relation
 \begin{equation}
\alpha = a/\lambda.
\end{equation} 
Figure~\ref{fig_2d_gf_2} presents the system size scaling of the total power and the power in the low-frequency component, corresponding to Fig.~\ref{fig_2d_gf_1}. Table~\ref{tab1} shows the numerically estimated critical exponents. We show the global fitness scaling function in Fig.~\ref{fig_gf_1}. The clean data collapse reasonably supports an accurate estimate of the critical exponents within the statistical error. The change in the spectral exponent $\alpha$ with dimensionality is evident, and the exponent tends to approach the mean-field value in dimension ${\rm D} \ge 4$.

\begin{figure}[t]
  \centering
       \scalebox{1.0}{\includegraphics{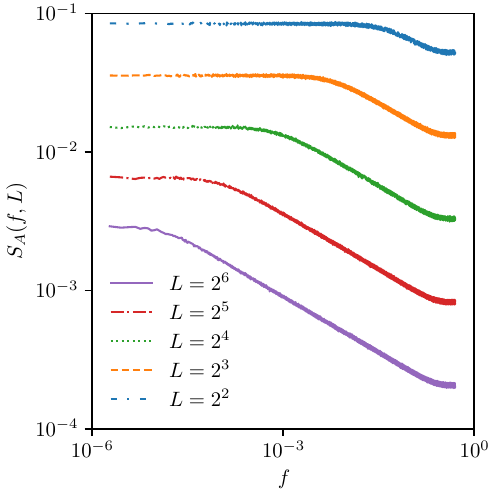}}
  \caption{The local activity power spectra for the 2D BS model with different system sizes. }
  \label{fig_2d_la_1}
\end{figure}

\begin{figure}[t]
  \centering
       \scalebox{1.0}{\includegraphics{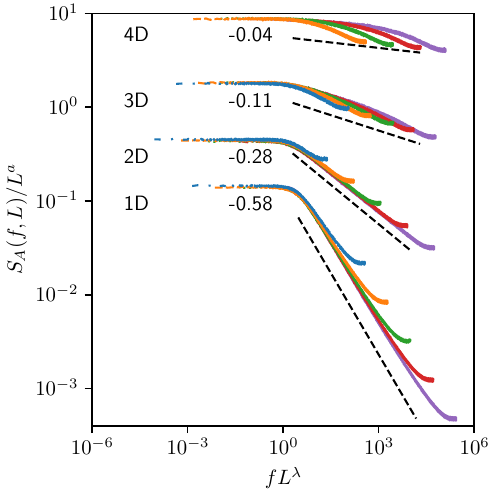}}
  \caption{The data collapse of the local activity power spectra for the BS model in different dimensions. }
  \label{fig_la_1}
\end{figure}

\subsection{Local activity}
Similarly, Fig.~\ref{fig_2d_la_1} shows the local activity power spectra for different system sizes in the 2D BS model. Although the power remains constant for the low-frequency regimes, a system size scaling emerges along with the $1/f^{\alpha}$ behavior for the non-trivial frequency regime $L^{-\lambda}\ll f \ll 1/2$. Interestingly, in the non-trivial frequency regime, the system size scaling of the local activity power spectra is of the form $1/L^{\rm D}$, implying [cf. Eq.~(\ref{eq_ps1})]  
\begin{equation}
a-\lambda\alpha = -{\rm D}.
\label{eq_bl}
\end{equation} 
In addition, we numerically note that the exponents satisfy a scaling relation valid below the upper critical dimension
\begin{equation}
\alpha+{\rm D}/\lambda = 1, ~~{\rm for}~~{\rm D}\le {\rm D}_u.
\label{eq_al}
\end{equation}  
Plugging $\alpha$ from Eq.~(\ref{eq_al}) into Eq.~(\ref{eq_bl}) yields $\lambda = a+2{\rm D}$. We show the local activity scaling function in Fig.~\ref{fig_la_1} for different dimensions. The estimated critical exponents are shown in Table~\ref{tab2}. Correspondingly, the spectral exponent $\alpha$ decreases with dimension and goes to zero (the mean-field value) in dimension ${\rm D}_u = 4$. 

In Ref.~\cite{PhysRevLett.73.2162}, it has been noted that $\alpha = 1-{\rm D}/\mathcal{D}$, where $\mathcal{D}$ is the avalanche dimension. One can immediately recognize that $\lambda = \mathcal{D}$. We infer that Eq.~(\ref{eq_al}) is an outcome of the fact that the underlying space-time scaling features are interlinked. At ${\rm D}_u = 4$, the value of the $\lambda$ seems consistent with the logarithmic correction as predicted for $\mathcal{D}$~\cite{PhysRevLett.84.2267}. Since $\alpha = 1-\tau_{\rm all}$ for ${\rm D}\le {\rm D}_u$~\cite{PhysRevE.53.414}, the estimated value of the spectral exponent agrees well with previously determined values of return time exponents in high dimensions~\cite{PhysRevLett.73.2162, PhysRevLett.80.5746}.

\begin{table}[t]
\centering
\begin{tabular}{|c|ccccc|}
\hline 

D & 1 & 2 & 3 & 4 & rBS in 2D \\		
\hline
$a$ & 3.02(3) & 4.85(2) & 6.55(4) & 8.13(8) & 4.00(1) \\
$\lambda$ & 2.43(4) & 3.04(3) & 3.66(5) & 4.2(1) & 2.07(2)\\
$\alpha = a/\lambda$ & 1.24(3) & 1.60(2) & 1.79(4) & 1.92(7) & 1.93(2)\\
\hline
\end{tabular}
 \caption{Numerically estimated critical exponents describe the global fitness power spectrum properties. The 2D rBS corresponds to mean-field behavior. }
\label{tab1}
\end{table}

\begin{table}[t]
\centering
\begin{tabular}{|c|ccccc|}
\hline 

D & 1 & 2 & 3 & 4 & rBS in 2D \\
\hline
$\lambda = a+2{\rm D}$ & 2.38(1) & 2.79(1) & 3.38(4) & 4.15(1) & 2.07(2)\\
$\alpha = 1-{\rm D}/\lambda$ & 0.58(1) & 0.28(1) & 0.11(2) & 0.04(1) & 0.03(1)\\
\hline
\end{tabular}
 \caption{The estimated critical exponents describe local activity power spectrum properties. }
\label{tab2}
\end{table}

 \section{Summary and Discussion}{\label{sec_5}} 
In summary, we studied the Bak-Sneppen evolution model on a hypercubic lattice in high dimensions. We compute power spectra for fitness noise and local activity with different system sizes to understand the temporal correlations. The power spectrum exhibits $1/f^{\alpha}$ noise with a \emph{nontrivial spectral exponent} in a frequency regime of $L^{-\lambda} \ll f \ll 1/2$. Employing finite-size scaling, we determine critical exponents and scaling functions. For fitness noise, the spectral exponent $\alpha$ increases with dimensionality from 1.2 in 1D to a mean-field value of 2 (Lorentzian spectrum) at the upper critical dimension ${\rm D}_u = 4$. For local activity, the spectral exponent $\alpha$ decreases with ${\rm D}$ and approaches a mean-field value of 0 (white noise) at ${\rm D}_u = 4$. In both cases, the ratio for the cutoff frequency exponent and dimension, i.e., $\lambda/{\rm D}$, does not differ appreciably, and this decreases with dimension and eventually approaches one at ${\rm D}_u = 4$.

Since the underlying space-time scaling features in the BS model are interlinked, the local activity power spectra reveal the links transparently. The avalanches are spatially fractal above the upper critical dimension while being compact for ${\rm D}\le {\rm D}_u$~\cite{PhysRevLett.84.2267}. Similarly, the local activity is recurrent for ${\rm D}\le {\rm D}_u$ and transient for ${\rm D}>{\rm D}_u$. To contrast, recall that the simple random walk remains recurrent below ${\rm D}\le 2$ and transient for ${\rm D}>2$~\cite{Krapivsky_2010}.
Notice that $\lambda/{\rm D} \to 1$ in the mean-field limit suggests the inverse of cutoff frequency as a function of the number of species grows linearly. However, below the upper critical dimension, the inverse of cutoff frequency varies in a \emph{nonlinear} manner (super-linearly), leading to a nontrivial spectral exponent for the $1/f^{\alpha}$ noise.

\section*{ACKNOWLEDGMENTS}
R.C. would like to acknowledge financial support through the Junior Research Fellowship at UGC, India. A.S. and A.C.Y. are supported by Banaras Hindu University through a fellowship (Grant No. R/Dev./Sch/UGC Non-Net Fello./2022-23/53315) and a seed grant under the IoE (Seed Grant-II/2022-23/48729), respectively.

\bibliography{s1sources}

\end{document}